\documentclass[11pt,a4paper]{article}
\pdfoutput=1
\usepackage{jheppub}	

\usepackage[normalem]{ulem}

\usepackage{graphicx}

\def\lsim{\raise0.3ex\hbox{$\;<$\kern-0.75em\raise-1.1ex
\hbox{$\sim\;$}}}
\def\gsim{\raise0.3ex\hbox{$\;>$\kern-0.75em\raise-1.1ex
\hbox{$\sim\;$}}}

\title{
Combining accelerator and reactor measurements of $\theta_{13}$:
the first result}
\author[a,b]{P.~A.~N.~Machado}
\author[c]{H.~Minakata}
\author[d]{H.~Nunokawa}
\author[a]{R.~Zukanovich Funchal}

\affiliation[a]{ Instituto de F\'{\i}sica, Universidade de S\~ao
  Paulo, C.\ P.\ 66.318, 05315-970 S\~ao Paulo, Brazil}
\affiliation[b]{ Institut de Physique Th\'eorique, CEA-Saclay, 91191
  Gif-sur-Yvette, France} \affiliation[c]{Department of Physics, Tokyo
  Metropolitan University, Hachioji, Tokyo 192-0397, Japan}
\affiliation[d]{ Departamento de F\'{\i}sica, Pontif{\'\i}cia
  Universidade Cat{\'o}lica do Rio de Janeiro, C. P. 38071, 22452-970,
  Rio de Janeiro, Brazil}

\emailAdd{accioly@fma.if.usp.br}
\emailAdd{hisakazu.minakata@gmail.com}
\emailAdd{nunokawa@puc-rio.br} 
\emailAdd{zukanov@if.usp.br}

\abstract{The lepton mixing angle $\theta_{13}$, the only
    unknown angle in the standard three-flavor neutrino mixing scheme,
    is finally measured by the recent reactor and accelerator neutrino
    experiments. We perform a combined analysis of the data coming
    from T2K, MINOS, Double Chooz, Daya Bay and RENO experiments and
    find $\sin^2 2\theta_{13}=0.096 \pm 0.013(\pm 0.040)$ at 1
    $\sigma$ (3 $\sigma$) CL and that the hypothesis $\theta_{13} = 0$
    is now rejected at a significance level of 7.7 $\sigma$.
    We also discuss the near future expectation on the precision of
    the $\theta_{13}$ determination by using expected data from these
    ongoing experiments.}

\keywords{Neutrino Physics, Standard Model}

\arxivnumber{1111.3330}

\begin{document}
\maketitle

\section{Introduction}

The accelerator search for $\nu_e$ appearance
\cite{Itow:2001ee}\footnote{For an updated version see 
  \href{http://neutrino.kek.jp/jhfnu/loi/loi.v2.030528.pdf}{{\tt 
http://neutrino.kek.jp/jhfnu/loi/loi.v2.030528.pdf}}} and the
precision measurement of reactor neutrino disappearance
\cite{Martemyanov:2002td,Minakata:2002jv} are both viable ways to
measure $\theta_{13}$, which has been, until very recently, the unique
unknown mixing angle of the lepton flavor mixing
matrix~\cite{Maki:1962mu}.  It must be stressed that the experimental
redundancy for measuring $\theta_{13}$ may be justified because of the
complementary nature of the two types of experiments, as discussed,
for example, in \cite{Minakata:2002jv,Hiraide:2006vh}.  While the
reactor experiments provide a clean measurement of $\theta_{13}$ which
is free from degeneracy
\cite{BurguetCastell:2001ez,Minakata:2001qm,Fogli:1996pv}, the
accelerator measurement can enjoy the interplay with the CP phase
$\delta_{\text{CP}}$, which connotes the possibility of extension of
the experiment to an upgraded phase to search for CP violation.

It is very fortunate to see that the era of simultaneous measurement
of $\theta_{13}$ by accelerator and reactor has just arrived.  In June
of 2011 the T2K group reported six clean events of $\nu_e$ appearance,
implying 2.5 $\sigma$ indication for non-zero $\theta_{13}$
\cite{Abe:2011sj} with a best fit value comparable to the CHOOZ limit
\cite{Apollonio:2002gd} (see also
\cite{Boehm:2001ik,Yamamoto:2006ty,Adamson:2010uj}).  It was soon
followed by the MINOS collaboration which reported also indication of
non-zero $\theta_{13}$ \cite{Adamson:2011qu}. At the end of 2011, one
of the reactor $\theta_{13}$ experiments, Double Chooz
\cite{Ardellier:2006mn}, reported their first result, constraining
$\theta_{13}$ to a range $\sin^22\theta_{13}$ = 0.086 $\pm 0.051 $ at
68 \% CL~\cite{Abe:2011fz,DC-1st}.\footnote{For the official Double
  Chooz results see
  \url{http://doublechooz.in2p3.fr/Status_and_News/status_and_news.php}.}

Very recently we were surprised by the announcement of another two
reactor $\theta_{13}$ experiments. First, Daya Bay\cite{Guo:2007ug},
which reported a measurement of $\theta_{13}$ as accurate as
$\sin^22\theta_{13}$ = 0.092 $\pm 0.017 $ at 68 \% CL. The
significance level for non-zero $\theta_{13}$ obtained by them is 5.2
$\sigma$~\cite{An:2012eh}.  Second, RENO\cite{Ahn:2010vy,RENO-LowNu},
which reported $\sin^2 2\theta_{13}$ = 0.113 $\pm 0.023 $ at 68 \% CL,
excluding a non-zero $\theta_{13}$ at 4.9 $\sigma$~\cite{RENO-first}.
Though still limited by both statistics and systematics (except for
Daya Bay whose systematics is already small), these results, together
with the aforementioned accelerator data, constitutes the most
valuable information on $\theta_{13}$ to date.  Therefore, we believe
that it is a meaningful step to attempt a combined
analysis of these data set.

The issue of possible non-zero $\theta_{13}$ has been discussed in the
context of global analyses which include the solar and the atmospheric
neutrino data even
before~\cite{Fogli:2008jx,Schwetz:2008er,GonzalezGarcia:2010er} or
after~\cite{Fogli:2011qn,Schwetz:2011zk} the T2K
result~\cite{Abe:2011sj}. However, given the current precision on the
determination of $\theta_{13}$, such global fits seem unnecessary in
this specific context. Hence, in this paper we restrict ourselves to a
combined analysis of the accelerator and the reactor $\theta_{13}$
experiments only.

\section{Analysis details}

We analyze the available accelerator data from T2K~\cite{Abe:2011sj}
and MINOS~\cite{Adamson:2011qu} in the $\nu_\mu \to \nu_e$ appearance
channel in combination with the very recent Double
Chooz~\cite{Abe:2011fz,DC-1st}, Daya Bay~\cite{An:2012eh} 
and RENO~\cite{RENO-first} reactor
data in the $\bar \nu_e \to \bar \nu_e$ disappearance channel.  
We will also make some prognostication to the near future.
The simulations were performed using a modified version of
GLoBES~\cite{Huber:2007ji}.

\subsection{Accelerator experiments: T2K and MINOS}

The T2K experiment uses a narrow 2.5$^\circ$ off-axis $\nu_\mu$ beam
generated at J-PARC in Tokai which is directed to the Super-Kamiokande
detector of fiducial mass 22.5 kt located in Kamioka 295 km away from
J-PARC.  In order to reproduce the T2K allowed region in the $\sin^2
2\theta_{13}$ - $\delta_{\rm CP}$ plane, reported in figure 6 of
ref.~\cite{Abe:2011sj}, we have simulated the T2K signal in the
$\nu_\mu \to \nu_e$ appearance channel in a similar way as done in
ref.~\cite{Hiraide:2006vh}.  We took the neutrino fluxes from the
letter of intent of the Hyper-Kamiokande project~\cite{Abe:2011ts} and
the background from \cite{Abe:2011sj}.  The cross sections and energy
dependent efficiencies for charged current quasi-elastic (CC-QE) and
non quasi-elastic (CC-NQE) events are simulated in a similar manner
as in \cite{Hiraide:2006vh} to reproduce the energy spectra given in
\cite{Abe:2011ts}.

Energy smearing and the consequent migration of events were taken into
account in our calculations by using a Gaussian energy resolution
function with width 85 (130) MeV for CC-QE (CC-NQE) events.  For
CC-NQE events, following the procedure described in the Appendix of
~\cite{Hiraide:2006vh}, a shift of 350 MeV was introduced in the
Gaussian smearing function in order to take into account the significant
difference between true and reconstructed neutrino energy.
In reproducing the current T2K result we assumed $1.43 \times 10^{20}$
POT and 23\% systematic uncertainly in the absolute normalization.

The MINOS experiment uses the NuMI beamline and operates with a near
detector located on-site at Fermilab, and a far detector located 735
km away in the Soudan Underground Laboratory.  The near (far) detector
consists of 0.98 kt (5.4 kt) of alternating layers of steel and
plastic scintillator. In order to reproduce the MINOS allowed region
in the $\sin^2 2\theta_{13} - \delta_{\rm CP}$ plane, given in figure 3 of
ref.~\cite{Adamson:2011qu}, we have simulated the $\nu_e$ signal using
the same procedure as in ref.~\cite{Kopp:2010qt} but with the
background and systematic uncertainties taken from
\cite{Adamson:2011qu}.  We assumed a total exposure of $8.2 \times
10^{20}$ POT, but a tuning of the normalization was needed in order to
obtain the correct number of signal events.

\subsection{Reactor experiments: Double Chooz, Daya Bay and RENO}

Double Chooz (DC) is a reactor antineutrino oscillation
experiment~\cite{Ardellier:2006mn} based on the CHOOZ-B Nuclear Power
Station.  The experiment is a double detector apparatus (each detector
with a fiducial volume of 10.3 m$^3$) based on liquid scintillator,
though until 2013 they will be taking data only with their far
detector located at 1.05 km from the two 4.27 GW$_{\rm th}$ reactor
cores.

To simulate the $\bar \nu_e \to \bar \nu_e$ disappearance reported by
DC collaboration in refs.~\cite{Abe:2011fz,DC-1st} we have performed a
calculation based on the far detector specification and reactor fuel
composition given in ref.~\cite{Ardellier:2006mn}, with systematic
uncertainties, background and efficiency, and other additional
information according to ~\cite{Abe:2011fz,DC-1st}.

Before analyzing the real data, we first tried to reproduce the expected 
visible energy spectra obtained by the Monte Carlo (MC) simulations of the 
DC collaboration, in the absence and presence of oscillation shown 
(respectively, by the blue dotted and red solid histograms) 
in figure 3 of ~\cite{Abe:2011fz}. 
Indeed, in our attempt to reproduce the visible energy spectra, 
we have noticed that these spectra exhibit significant {\em distortions}  
if compared to the corresponding spectra as a function of the true 
prompt energy, which, of course, can not be measured directly. 

In order to mimic such a rather strong {\em distortions}, 
which are due to various effects taken into account in the MC simulations
by the DC collaboration, we first introduce an energy smearing effects using a
Gaussian energy resolution function with a width $\sigma_E =
12\%\sqrt{(E/\text{MeV})} + 0.15 $ MeV.  We note that due to the 2nd
term in $\sigma_E$, we can reproduce rather well the spectra after
taking into account the additional corrections described below.  We,
however, stress that the inclusion or omission of the 2nd term in
$\sigma_E$ does not alter much the allowed parameter region of
$\sin^2 2\theta_{13}$ and $\delta_{\text{CP}}$ presented in this paper, 
though it affects the $\chi^2_{\text{min}}$ values.

In addition to the energy smearing we have further taken into account,
in an approximate way, two kinds of corrections which were actually
introduced in the analysis by the DC collaboration~\cite{DC-corrections}\footnote{Relevant information to be available at 
\url{http://doublechooz.in2p3.fr/Status_and_News/status_and_news.php}. }
in order to understand their data.  The first one is a non linearity
correction. This is based on the energy calibration by using several
sources performed by the DC collaboration.  Roughly speaking, the
observed visible energies (or to be more precise the number of
photoelectrons) tend to be overestimated (underestimated) for energy
larger (smaller) than $\sim 1.5 $ MeV for up to a few percent, when
compared to the ones predicted by MC simulations.  Note that the
correction is energy dependent, see ~\cite{DC-corrections}.  The
second correction is the one based on the $Z$-dependence calibration,
which shows that the observed energy tends to be underestimated when
the neutrino event occurs in the region far from the center of the
detector, for up to a few percent~\cite{DC-corrections}.  We note
that, after taking into account these two corrections in addition
to the energy smearing, we can reproduce reasonably well the energy
spectra shown in figure 3 of ~\cite{Abe:2011fz}.

Daya Bay (DB) experiment measures $\bar \nu_e$ from six 2.9 GW$_{\rm
  th}$ reactors grouped into three pairs of nuclear power plants (NPP)
using six detectors deployed in a near-far arrangement allowing to
compare rates at various baselines.  There are two near detectors at
364 m from the Daya Bay NPP, one near detector at 480 m (528 m) from
Ling Ao (Ling Ao-II) NPP and three far detectors at 1912 m (1540 m)
from Daya Bay (Ling Ao and Ling Ao-II) NPP~\cite{An:2012eh}. Each one
of these identical detectors is made of a 5 m diameter cylindrical
stainless steel vessel which holds a 3.1 m diameter inner vessel and a
4 m diameter outer vessel.  The inner vessel holds 20 ton of
Gadolinium-doped liquid scintillator (the target) which is shielded
from the 20 ton liquid scintillator in the outer acrylic vessel by 37
ton of mineral oil. We have implemented in our simulation the
systematic uncertainties and the background in accordance with
\cite{An:2012eh}.  In this paper, for DB, we do not consider the
energy spectrum but restrict ourselves to the rates only analysis, as
done in ref.~\cite{Guo:2007ug}, even for our near future analysis to
be discussed in section~\ref{sec:future}. The reason is that DB's
systematic uncertainty for the rate only analysis is already quite
small and their data are still statistically limited and hence it
seems that the spectrum information would not play an important role
for the time being.

RENO is the reactor experiment which receives neutrinos from the
YongGwang Nuclear Power Plant located 400 km from Seoul in which six
2.8 GW$_{\rm th}$ reactors are lined up.  They use two 16 t liquid
scintillator identical detectors, the near (far) detector located at
roughtly 300 m (1.3 km) from the reactors.  RENO has been taking data
with both detectors since August 2011.  In order to simulate RENO
$\bar \nu_e$ disappearance signal we performed rate-only analysis,
using the background, energy resolution and the systematic
uncertainties given in ref.~\cite{RENO-LowNu, RENO-first}.

In DC, DB and RENO simulations we have used the new reactor
antineutrino flux calculations~\cite{Mueller:2011nm,Huber:2011wv}.
This has little impact on our results, since DC is normalized to
Bugey-4 cross section and the other reactor experiments have near
detectors.

\begin{figure*}[!t]
\begin{center}
 \includegraphics[width=0.49\textwidth]{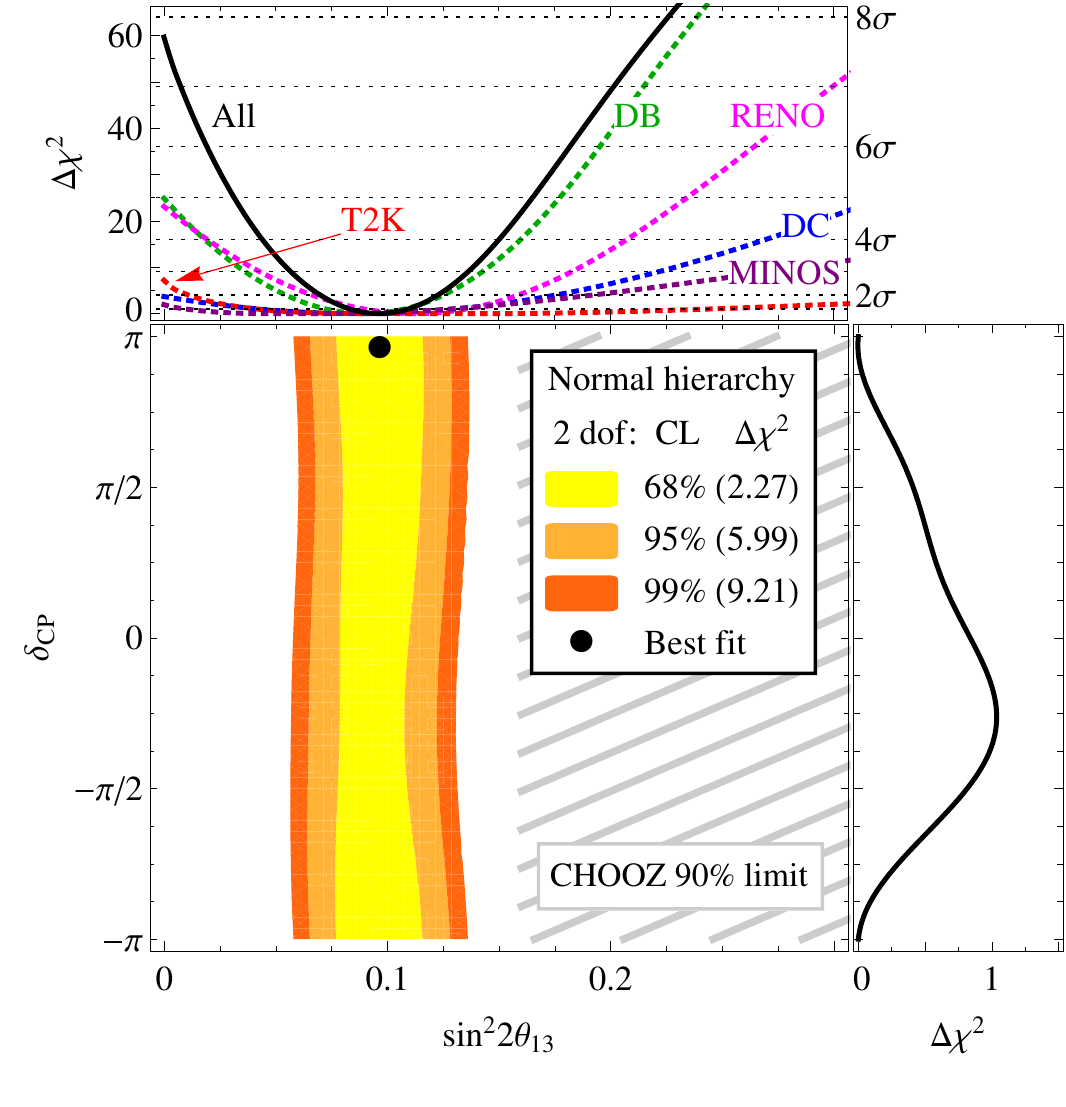}
 \includegraphics[width=0.49\textwidth]{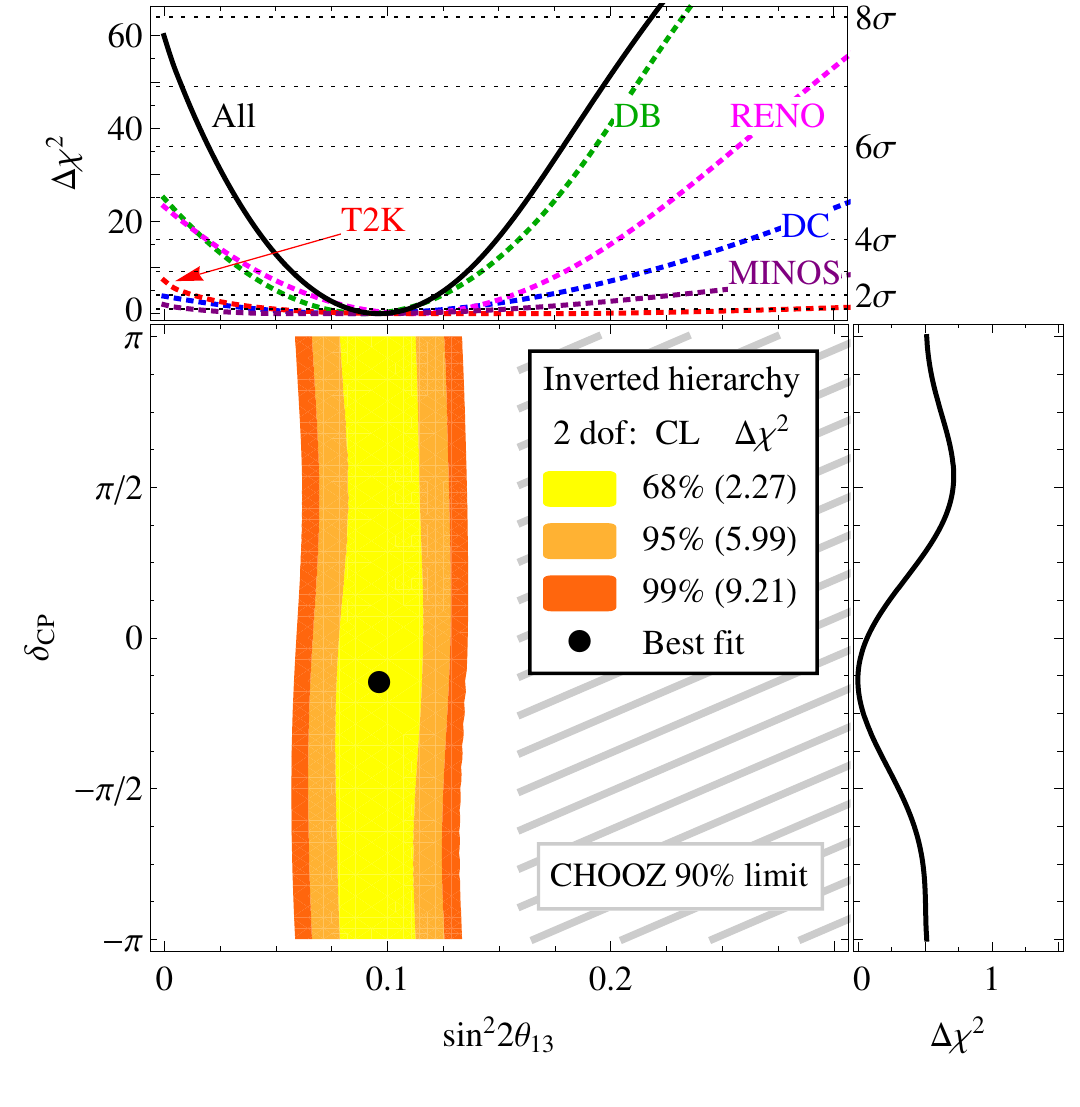}
\vspace{-2mm}
\end{center}
\vspace{-8mm}
\caption{Allowed region in $\sin^2 2\theta_{13} - \delta_{\rm CP}$
  plane for T2K, MINOS, Double Chooz (DC), Daya Bay (DB) and RENO
  combined at 68\%, 95 \% and 99\% CL for 2 dof, assuming normal (left
  panel) or inverted (right panel) mass hierarchy.  We also show the
  $\Delta \chi^2$ behavior as a function of $\sin^2 2\theta_{13}$
  (top) and as a function of $\delta_{\rm CP}$ (right) in each
  case. As a reference we also show the 90\%CL exclusion limit from
  CHOOZ~\cite{Apollonio:2002gd}.  }
\label{combination}
\end{figure*}

\section{Analysis results: current status}

\subsection{Combining accelerator and reactor data}

Before combining the accelerator and reactor neutrino data we have
verified that we are able to reproduce quite well the individual
result of each experiment T2K~\cite{Abe:2011sj},
MINOS~\cite{Adamson:2011qu}, DC~\cite{DC-1st}, DB~\cite{An:2012eh} 
and RENO~\cite{RENO-first}.
Here we present our combined analysis of these experiments.

In figure~\ref{combination} we show the allowed region obtained in our
combined analysis. The yellow, orange, and red bands correspond,
respectively, to 68\%, 95\%, 99\% CL regions for 2 degree of freedom
(dof). We also show the behavior of $\Delta \chi^2$ for 1 dof as a
function of $\sin^2 2\theta_{13}$ (attached to top), and as a function
of $\delta_{\rm CP}$ (attached to right side) in each panel.
For T2K, DB  and RENO, only the total rate was considered,
whereas for MINOS (DC) we used data of 7 (18) energy bins. However,
we have checked that T2K allowed region does not change much if we
also take into account the spectrum information.
In our fit we have explicitly assumed one of the mass hierarchies
(normal or inverted) as input and varied $\sin^2 2\theta_{23}$ and
$\vert\Delta m^2_{32}\vert$, imposing Gaussian priors based on the
atmospheric neutrino experiments~\cite{Wendell:2010md} and
MINOS~\cite{Adamson:2011ig} results.  We observe that if we combine
only T2K and DC (not shown in figure~\ref{combination}), our allowed
regions agree very well with the result shown in \cite{DC-1st} for the
same fixed values of $\sin^2 2\theta_{23}$ and $\vert\Delta
m^2_{32}\vert$.

We conclude that at 95\% CL, the allowed range for $\theta_{13}$ is
given as $ 0.070 < \sin^2 2\theta_{13} < 0.122$ irrespectively of the 
mass hierarchy for 1 dof.  In the case of normal
(inverted) mass hierarchy, the best fit point is given by $\sin^2
2\theta_{13}=0.096$ ($\sin^2 2\theta_{13}=0.096$) and $\delta_{\rm CP}
= 0.97 \pi$ ($\delta_{\rm CP} = -0.14 \pi$) which correspond to
$\chi^2_{\rm min}/(24-2) =1.57$ (1.55).  At the moment, there is not
much significance in the preferred value of $\delta_{\rm CP}$.  We can
also see the contribution of each individual experiment to the
determination of $\sin^2 2\theta_{13}$. Currently  DB is the most powerful 
experiment in constraining $\sin^2 2\theta_{13}$ from both ends. Before DB 
and RENO announced their results T2K was the most effective experiment  
in excluding a vanishing value of $\sin^2 2\theta_{13}$, but
allowed for higher values of $\sin^2 2\theta_{13}$ than MINOS and DC.
The combination of the five experiments can now exclude $\sin^2
2\theta_{13}=0$ at 7.7  $\sigma$ CL, irrespectively of the mass
hierarchy.

\subsection{Potential hint on CP violation}

It was proposed in \cite{Minakata:2003wq} that hints of CP violation
could be obtained by combining accelerator and reactor
measurements. In this method, determining $\text{sgn}(\sin
\delta_{\text{CP}}$) is essentially the goal to reach.  At this
moment, however, change in $\Delta \chi^2$ is quite mild, $\lsim 1.0$,
as $\delta_{\text{CP}}$ is varied, as we can see from
figure~\ref{combination}.  Clearly, it is not possible to make any
definitive statements about which sign of $\sin \delta_{\text{CP}}$ is
preferred. Nevertheless, we may say that the region $\text{sgn}(\sin
\delta_{\text{CP}})>0$ $(\text{sgn}(\sin \delta_{\text{CP}})<0)$ is
slightly preferred in the normal (inverted) mass hierarchy case. This
tendency could become clearer by future accumulation of the data, as
shown in section \ref{sec:future}.

We note that another hint of $\text{sgn}(\sin \delta_{\text{CP}})$ is
provided by the three flavor analysis of the SK atmospheric neutrino
data~\cite{Takeuchi:2011aa,Kajita:2011zz}, which indicates, though
mildly, the negative $\sin \delta_{\text{CP}}$ region, which is more
prominent in the inverted hierarchy case.  We believe that the issue
of preferred region of $\delta_{\text{CP}}$ deserves careful watching with
accumulation of various experimental data in the future.

\section{Expectation: one year from now}
\label{sec:future}

We now make some predictions for the possible situation of
$\theta_{13}$ in the near future, about one year from now.  For
definiteness, in our simulations for the future expectation, we assume
the true parameters to be our best fit value $\sin^2 2\theta_{13}
  = 0.096$, $\delta_{\text{CP}} = 0.97\pi$ and the normal hierarchy scheme,
though we confirmed that the results do not change much even if the
inverted hierarchy (with the corresponding best fit values) was
assumed.  We do not consider MINOS in our predictions because the
impact of the improvement of MINOS sensitivity to $\theta_{13}$
appears to be limited.  We include the energy spectrum information in
the analysis of T2K.  We used the same priors as before for $\vert
\Delta m^2_{32}\vert$ and $\sin^2 2\theta_{23}$.  While this may seem
too conservative, these uncertainties mainly have an effect on the
upper bound on $\sin^2 2\theta_{13}$.

We take the same systematic uncertainties and the backgrounds claimed
by the experiments, as in the previous section. (For T2K and DC, we
also considered the case with reduced systematic uncertainties, see
the footnote \ref{footnote:sys} and text below.)  
We assume that DC have been taking data since April 2011 with averaged
77.5\% data taking efficiency for physics and 76\% reactor power
efficiency to take into account reactor off periods. For RENO, we set
the data taking to start at August 2011 and used the efficiencies and
DAQ live times (proportionally) given in ref.~\cite{Ahn:2010vy}.

We assume T2K will resume its operation in January 2012 with
their proposed integrated luminosity of $10^{21}$ POT/year.\footnote{
We know that this assumption no longer holds. However, we remain with
it because we do not know for sure the real situation of the T2K
experiment in 2012.  Therefore, as far as T2K is concerned, our
predictions can be viewed as optimistic. }
For concreteness, we assume the present configuration of DB throughout the
year. As they increase the number of detectors in the far experimental
hall one can shift our results by the appropriate number of
days/months to roughly account for that.

\begin{figure*}[!t]
\begin{center}
 \includegraphics[width=0.45\textwidth]{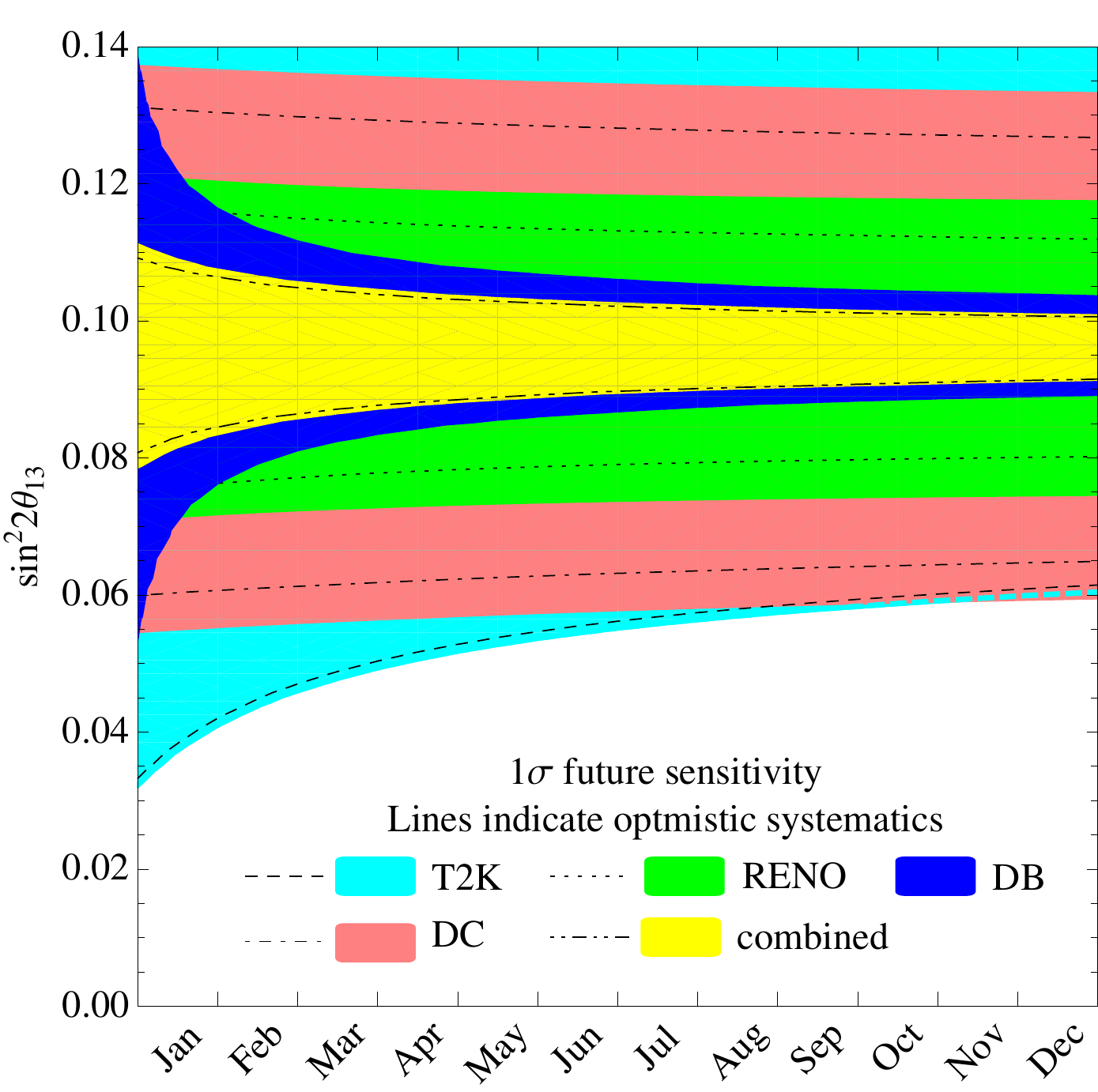}
 \includegraphics[width=0.49\textwidth]{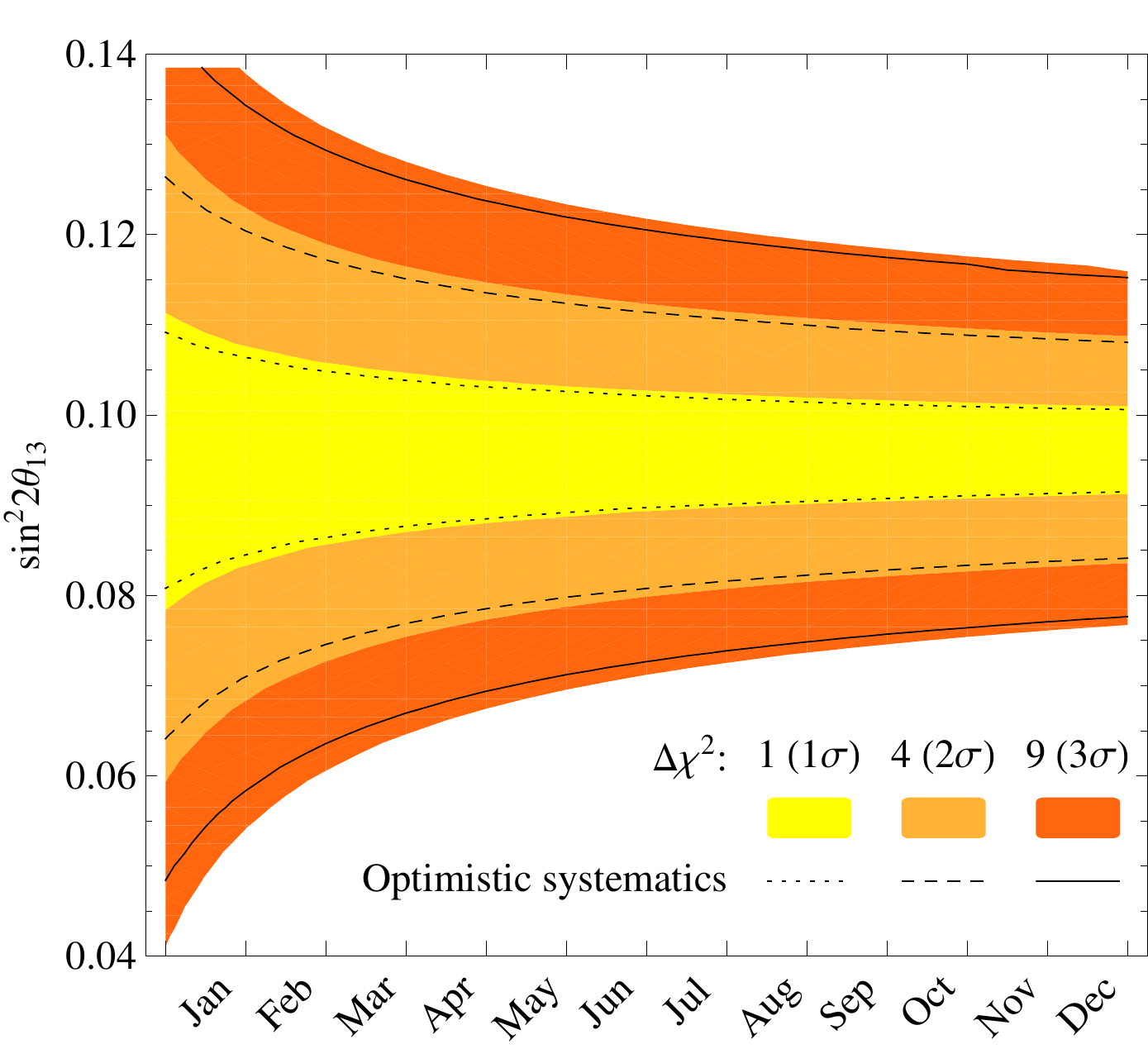}
\vspace{-2mm}
\end{center}
\vspace{-4mm}
\caption{ On the left panel, we show the expected 1 $\sigma$
  uncertainty on $\sin^2 2\theta_{13}$ for the case where the true
  value of $\sin^2 2\theta_{13} = 0.096$ and $\delta = 0.97\pi$
  (current best fit for the normal hierarchy) as a function of the
  months in 2012 for DC, RENO, T2K, DB as well as the combined case.
  On the right panel, we show the expected 1-3 $\sigma$ uncertainties
  on $\sin^2 2\theta_{13}$ as a function of time for the same input
  but only for the combined case.  On the left panel we indicate the
  effect of the improved systematics considered for T2K and DC by the
  dashed and dash-dotted lines, respectively, whereas their impact on
  the combined analysis is indicated by the dotted line on the left
  panel and by the dotted, dashed and solid lines on the right panel
  (see legends in the plots).  In fitting, the hierarchy is assumed to
  be unknown.  }
\label{fig:future-best}
\end{figure*}

On the left panel of figure~\ref{fig:future-best} we show the expected 1
$\sigma$ uncertainty on the determination of $\sin^22\theta_{13}$ as a
function of time, for the different experiments. We employ the
following color code for the bands: pink for DC, green for RENO, light
blue for T2K, blue for DB and yellow  for the combination.

We observe that, at this moment, DB with six detectors is the most
powerful experiment among the four and dominates the final combined
result.  RENO, the next powerful one, would have dominated the
combination at the very beginning of this year. However, as soon as
their result becomes dominated by systematic uncertainties they cannot
improve their sensitivity much. To do that they will have to improve
their systematics and/or do a spectrum analysis. DB, due to its high
reactor power, overall detector mass and smaller systematic
uncertainty, quickly becomes the most powerful experiment and remains
so throughout the year.  DC sensitivity can not improve much with only
the far detector. Regarding the bound on $\sin^2 2\theta_{13}$ from
below in 2012, T2K is comparable to DC, however both will not reach
the discriminability of RENO and DB.

We note that although a reduction of systematic uncertainties by
$\sim$ 30-50\% for T2K and DC~\footnote{
\label{footnote:sys}
For the DC experiment we use for each systematic uncertainty the best
value between the one quoted in their proposal and the one presented
in their latest paper~\cite{Abe:2011fz}. We also apply a 30\%
reduction of the number of their background events.  For T2K we assume
about 50\% reduction of systematic uncertainties, arbitrarily
re-scaling the current normalization uncertainty to 10\%.  For RENO,
we only contemplate the possibility of reducing the backgrounds
by 40\% at the cost of decreasing the near and far detector signal in
10\% and 3\%, respectively, while for DB we do not consider any
reduction of the systematics due to the already extraordinarily
 low systematics of this experiment.}
have some impact on the individual results of these experiments, it
does not essentially affect the accuracy of determination of
$\theta_{13}$ based on the combined data in the present
circumstances. On the other hand, a possible reduction of RENO
  backgrounds can fairly affect the final sensitivity to
  $\theta_{13}$. See figure~\ref{fig:future-best}. The improvement in
T2K is less visible probably because it is still dominated by
statistical error.

On the right panel of figure~\ref{fig:future-best}, we show the
$1\sigma-3\sigma$ uncertainty regions for the determination of
$\sin^22\theta_{13}$ as a function of time for all experiments
combined.  Here the yellow, orange and red bands correspond,
respectively, to $1\sigma$, $2\sigma$, and $3\sigma$ regions.
From this analysis we conclude that within 1 year, the uncertainty on
the determination of $\sin^22\theta_{13}$ at 1$\sigma$ CL may be
reduced from 0.013 to $\sim$ 0.005.  We have verified that this is also true
for the case where the true mass hierarchy is the inverted one.

At the same time, the hypothesis of a vanishing $\sin^22\theta_{13}$
could be rejected at a level of very high significance. We have
verified, under the above stated assumptions, that by the middle (end)
of 2012, the $\sin^2 2\theta_{13}=0$ hypothesis could be rejected with
a significance larger than 11 $\sigma$ (14 $\sigma$), if the future
data is consistent with the current best fit point.
We confirm that the impact of the improvement of systematics for T2K
and DC on the combined analysis is quite small also for the 2 and 3
$\sigma$ regions (see dashed and solid lines on the right panel of
figure~\ref{fig:future-best}).

\begin{figure*}[!t]
\begin{center} 
\includegraphics[width=0.49\textwidth]{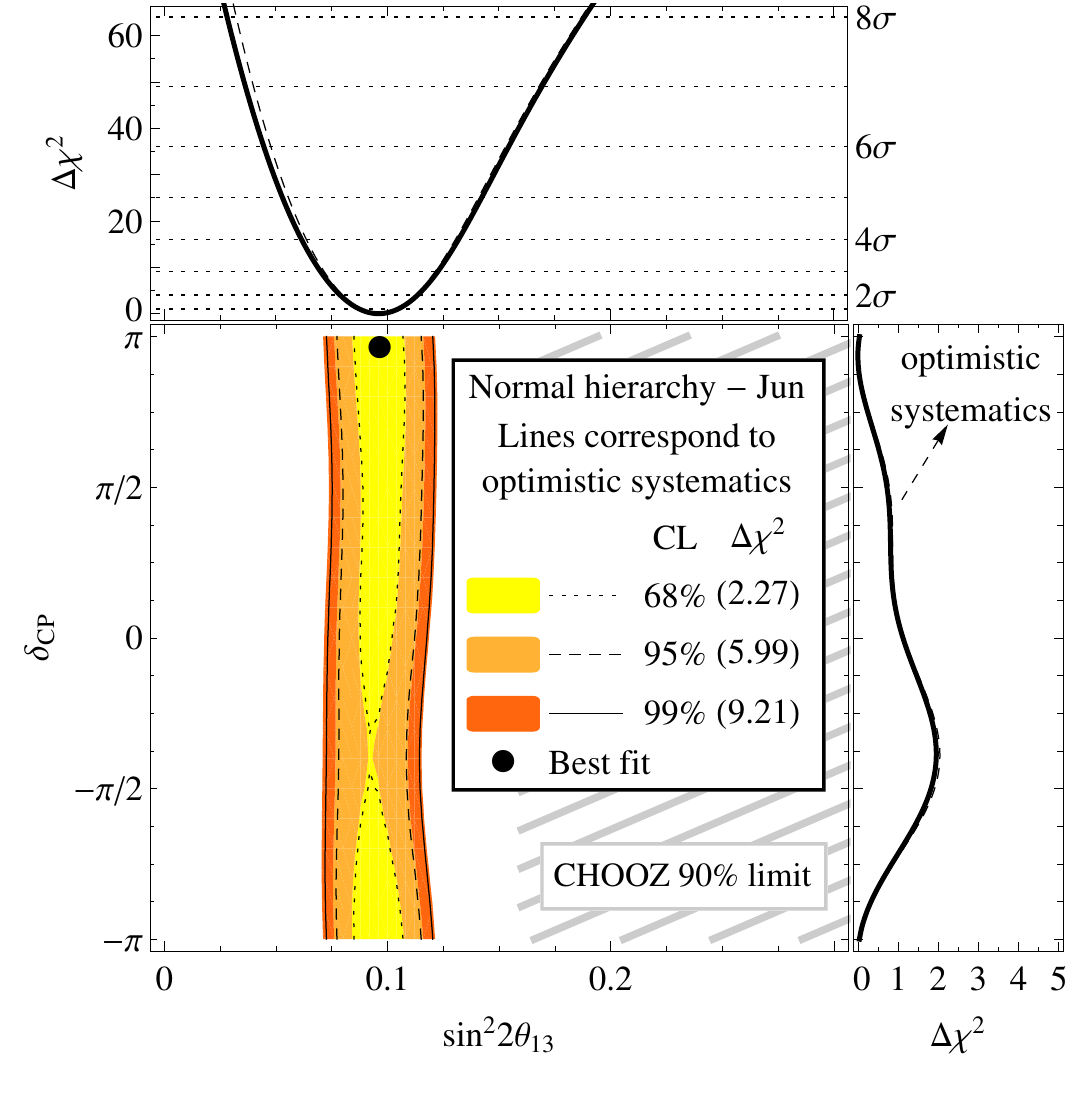}
\includegraphics[width=0.49\textwidth]{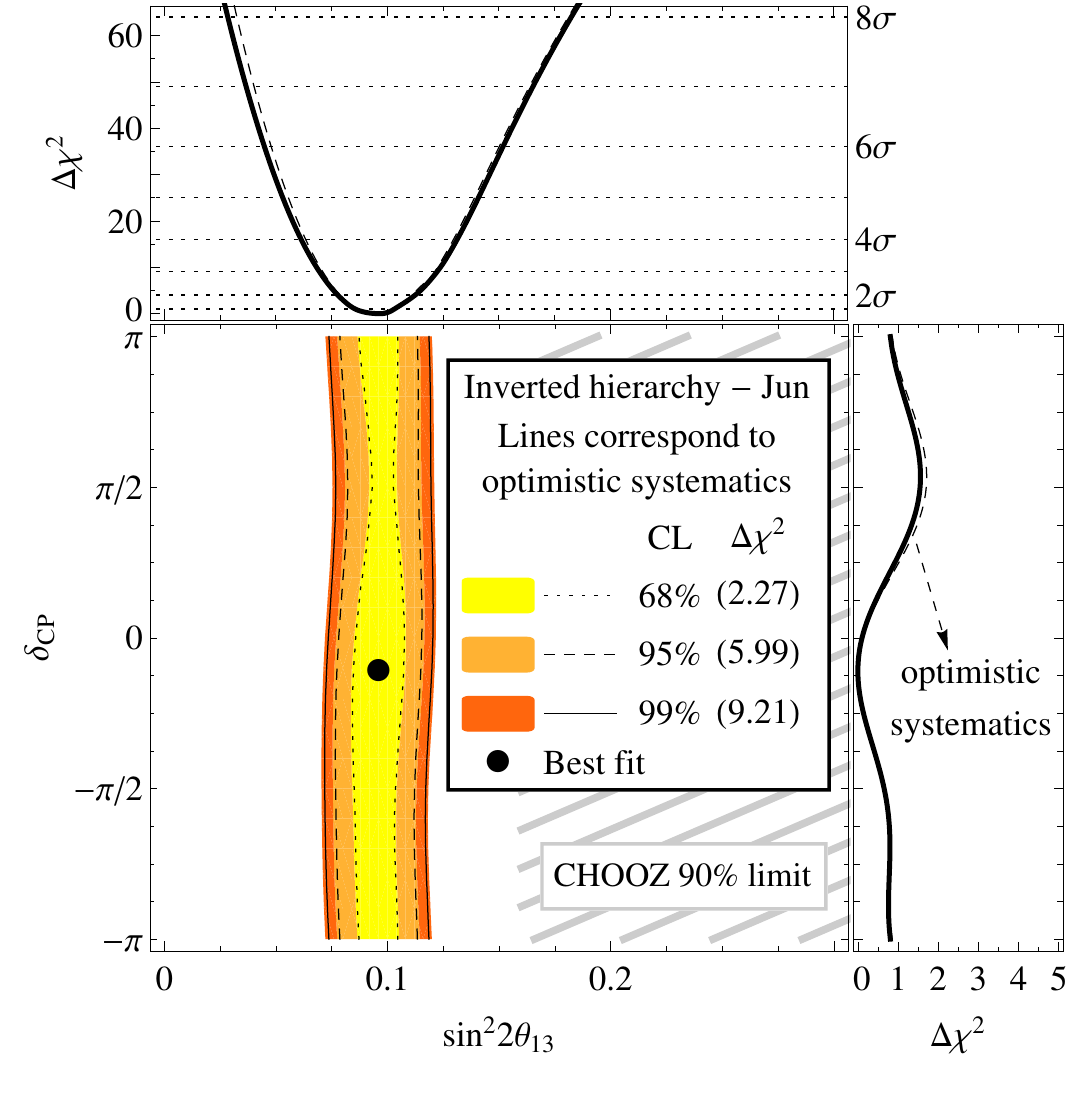}
\vspace{-2mm}
\end{center}
\vspace{-8mm}
\caption{Predicted allowed region in the 
$\sin^2 2\theta_{13}-\delta_{\rm CP}$ plane for T2K, MINOS, DC, DB and RENO
  combined at 68\%, 95 \% and 99\% CL for 2 dof in the middle (June) 
  of 2012, assuming normal (left panel) or inverted (right panel) mass
  hierarchy and as input the normal hierarchy best fit point of our
  current analysis.}
\label{fig:2012-june}
\end{figure*}

\begin{figure*}[!t]
\begin{center} 
\includegraphics[width=0.49\textwidth]{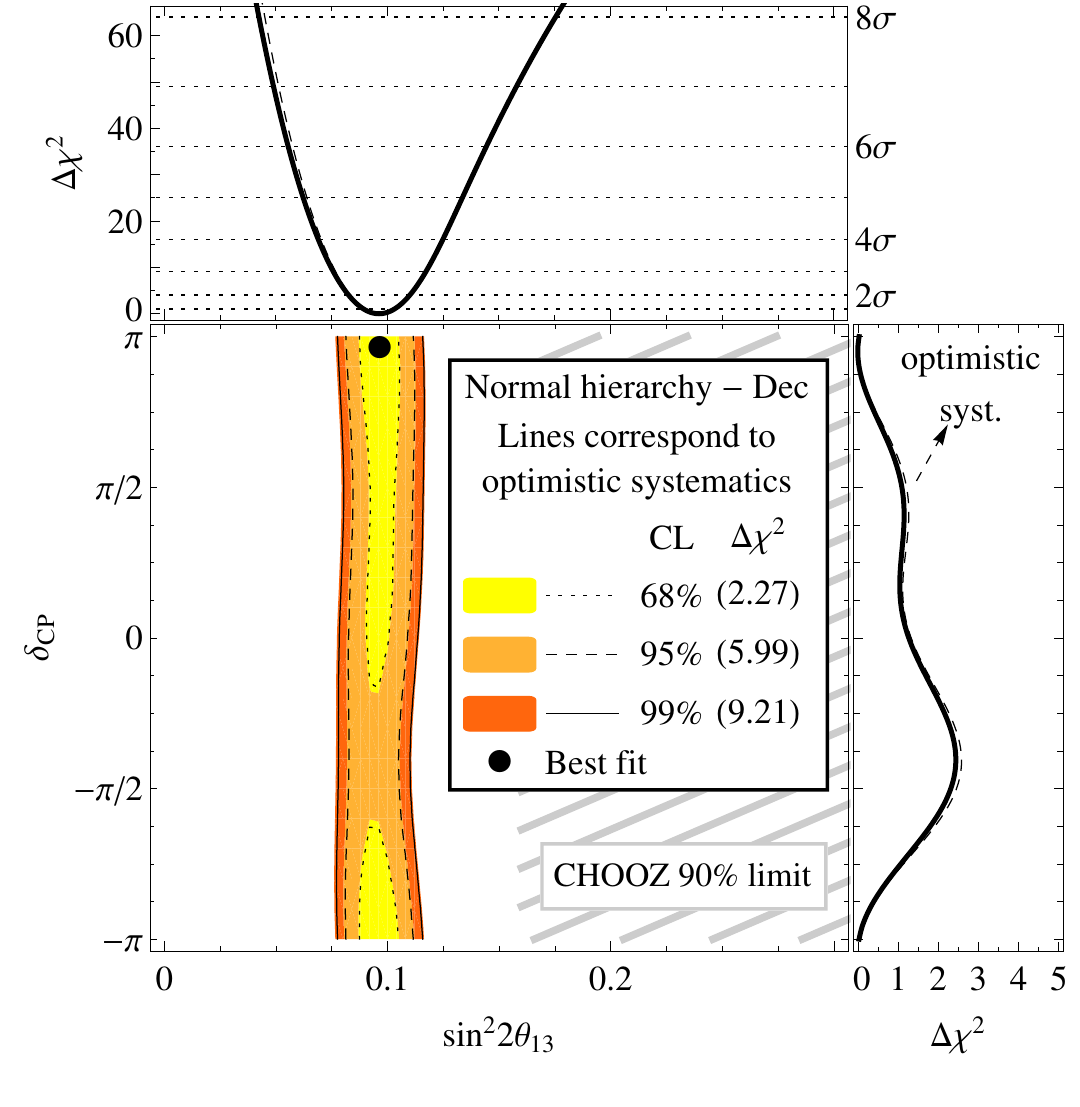}
\includegraphics[width=0.49\textwidth]{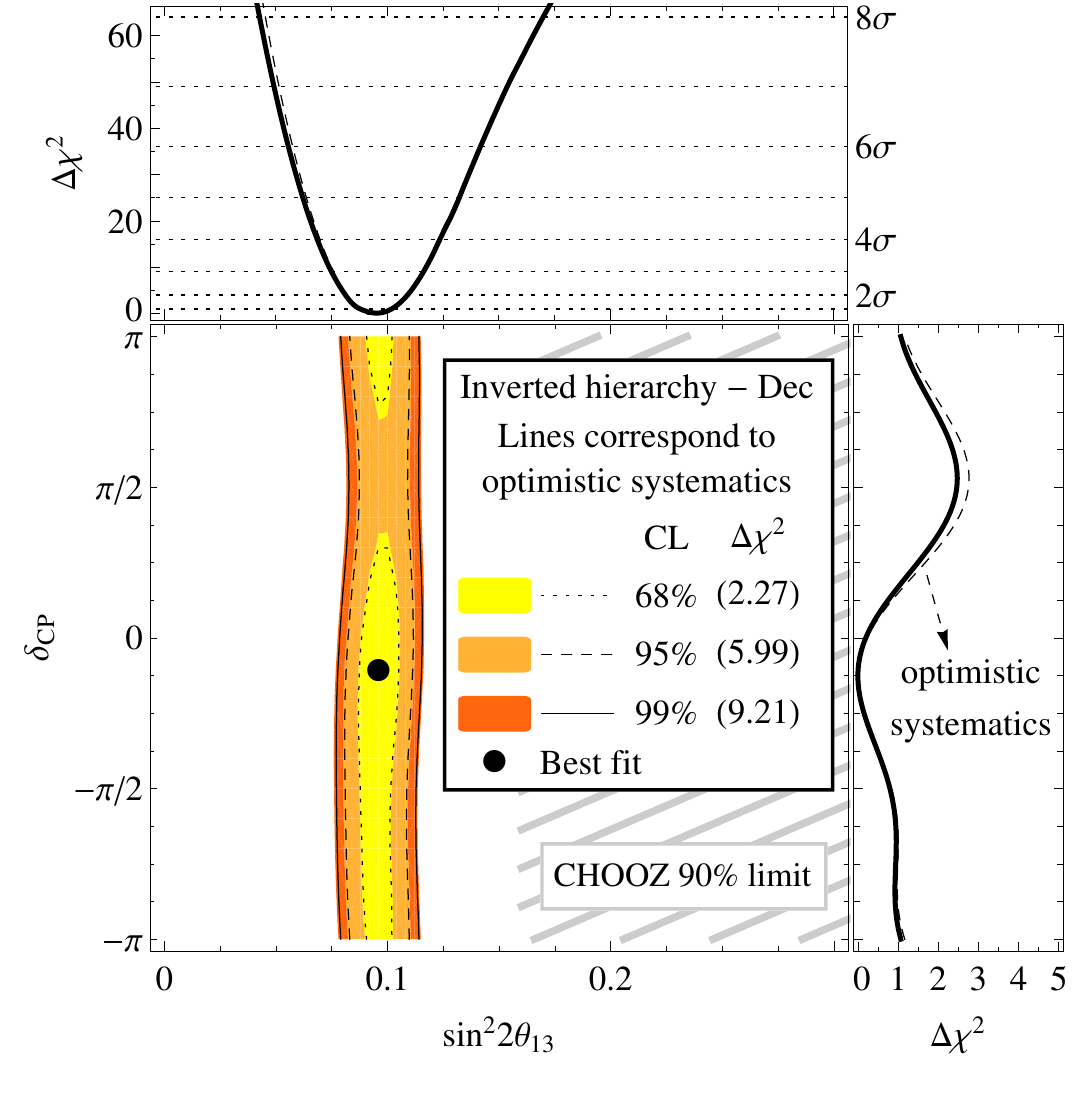}
\vspace{-2mm}
\end{center}
\vspace{-8mm}
\caption{Same as in figure~\ref{fig:2012-june} but 
for the end of the year (December 2012).}
\label{fig:2012-december}
\end{figure*}

In figures~\ref{fig:2012-june} and ~\ref{fig:2012-december}, using the
same format as in figure~\ref{combination}, we show the allowed region
in the $\sin^2 2\theta_{13} - \delta_{\rm CP}$ plane expected in June
and in December of 2012, respectively, that could be achieved by
combining T2K, MINOS, DC, DB and RENO data.  As input, we used the
best fit point for the normal hierarchy and fitted for each normal and
inverted mass hierarchy. As expected, from the right panel of
figure~\ref{fig:future-best}, the impact of the reduction of the
systematic uncertainties we considered in this work on the
determination of the parameter regions, as well as in the behavior of
$\Delta \chi^2$ (indicated by the dashed lines in
figures~\ref{fig:2012-june} and ~\ref{fig:2012-december}), is quite
small as far as the results expected in the near future ($\sim $ 1
year) are concerned.

Finally, we note that at the end of this year the combined $\Delta
\chi^2$ for different values of $\delta_{\text{CP}}$ is expected to be $\sim$
1-4, depending on the fitted hierarchy. This might be used as a hint
on which region of $\delta_{\text{CP}}$ is preferred, but this still will not
be strong enough to definitively pin down the value of $\delta_{\text{CP}}$
with high significance.  For future prospects on the
reactor-accelerator combined method, see also \cite{Huber:2009cw}.

\section{Conclusion}

We performed a combined analysis of the currently available
accelerator and reactor data which provide a very significant
evidence of non-zero $\theta_{13}$.  Being outside of the experimental
collaborations our simulation may be incomplete by lack of sufficient
information on backgrounds, systematic uncertainties, efficiencies,
etc.  However, we believe that we did a reasonable job and our results
serve as an independent confirmation of the analyses provided by the
experimental groups.

It is encouraging to see that the confidence level for non-zero
$\sin^2 2\theta_{13}$ now reaches $\simeq 7.7\ \sigma$ thanks, in
particular, to DB and RENO, in addition to T2K, MINOS and DC
experiments.  Still in this year, we will have indisputable evidence
for non zero $\theta_{13}$.
We predict that if the future data continues to be compatible with the
current best fit value $\sin^2 2\theta_{13}=0.096$, by the middle
(end) of 2012 $\sin^2 2\theta_{13}$ will be known within $\pm 0.007$
($\pm 0.005$) at 68\% CL.
Finally, we also studied the impact of the possible reduction of the
systematic uncertainties for DC (by 30\%) and T2K (roughly by half),
as well as for RENO backgrounds (by 40\%) while forfeiting signal
efficiency (3\% in the far detector). We have, however, found that the
reduction of errors do not affect in a significant way the combined
sensitivity at the end of 2012.

\begin{acknowledgments}
We thank Masashi Yokoyama for his generous help which enabled us to
have a better understanding of the T2K and HK simulations. We are
grateful to Tsunayuki Matsubara, Junpei Maeda, and Soo-Bong Kim for
useful discussions on some details of the results and analyses of the
DC and RENO experiments.
This work was supported by Funda\c{c}\~ao de Amparo \`a Pesquisa do
Estado de S\~ao Paulo (FAPESP), Funda\c{c}\~ao de Amparo \`a Pesquisa
do Estado do Rio de Janeiro (FAPERJ), Conselho Nacional de Ci\^encia e
Tecnologia (CNPq) and by the European Commission under the contracts
PITN-GA-2009-237920.  One of the authors (H.M.) is grateful to FAPERJ,
for its support which enabled his visit to the Departamento de
F\'{\i}sica, Pontif{\'\i}cia Universidade Cat{\'o}lica do Rio de
Janeiro, where part of this work was performed.  H.M. is supported in
part by KAKENHI, Grant-in-Aid for Scientific Research No. 23540315,
Japan Society for the Promotion of Science.
\end{acknowledgments}

\bibliographystyle{JHEP}
\bibliography{./accelerator-reactor-JHEP}

\end{document}